\documentclass[conference]{IEEEtran}
\IEEEoverridecommandlockouts

\usepackage{cite}
\usepackage{amsmath,amssymb,amsfonts}
\usepackage{algorithmic}
\usepackage{graphicx}
\usepackage{textcomp}
\usepackage{xcolor}
\usepackage{booktabs}
\usepackage{array}
\usepackage{tabularx}
\usepackage{multirow}
\usepackage{url}
\usepackage{hyperref}

\hypersetup{
    colorlinks=true,
    linkcolor=blue,
    citecolor=blue,
    urlcolor=blue
}

\def\BibTeX{{\rm B\kern-.05em{\sc i\kern-.025em b}\kern-.08em
    T\kern-.1667em\lower.7ex\hbox{E}\kern-.125emX}}

\begin{document}

\title{Designing a GDPR-Compliant Security Architecture for Remote Elderly Care Systems: A Privacy-by-Design Approach}

\author{
\IEEEauthorblockN{Md. Rahid Parvez}
\IEEEauthorblockA{
School of ICT and Industrial Management\\
Metropolia University of Applied Sciences\\
Helsinki, Finland\\
rahid.parvez@metropolia.fi}
\and
\IEEEauthorblockN{Mikael Soini}
\IEEEauthorblockA{
School of ICT and Industrial Management\\
Metropolia University of Applied Sciences\\
Helsinki, Finland\\
mikael.soini@metropolia.fi}
}

\maketitle

\footnotetext{Md. Rahid Parvez is with the School of ICT and Industrial Management at Metropolia University of Applied Sciences, Helsinki, Finland (e-mail: rahidparvez007@gmail.com). This work was completed as part of a Master's thesis at Metropolia University of Applied Sciences, April 2026.}

\begin{abstract}
IoMT-based remote elderly care systems generate continuous streams of sensitive health data, yet existing security architectures have not simultaneously addressed three interdependent challenges: GDPR-compliant edge-layer pseudonymisation, elderly-specific zero-interaction usability as a binding architectural constraint, and integrated STRIDE-based threat validation within a single unified design. This paper presents the Secure Edge Gateway (SEG) framework - a software-simulation-validated integrated IoMT security 
architecture for elderly care designed to resolve all three 
dimensions of this tripartite gap simultaneously. An ESP32-WROOM-32 residential gateway enforces MAC address whitelisting, HMAC-SHA256 cryptographic pseudonymisation before any network transmission, AES-128-CBC payload encryption, and TLS 1.3 transport security, in compliance with GDPR Articles 25 and 32. The framework is validated through software-based simulation, full STRIDE threat modelling across all six categories, attack tree analysis, GDPR compliance mapping across nine regulatory obligations, and a Data Protection Impact Assessment (DPIA) under Article 35. Published benchmarks confirm MQTT consumes 6--8\% less energy than HTTP in comparable IoT deployments, and edge processing achieves sub-50~ms response latency versus 200--700~ms for cloud-only systems. The results demonstrate that GDPR compliance and operational efficiency are complementary -- not competing -- objectives in resource-constrained IoMT deployments for elderly care.
\end{abstract}


\footnotetext{A prior version of this work was published as a Master's thesis in the Theseus Open Repository of Metropolia University of Applied Sciences (URN:NBN:fi:amk-202604186859), April 2026.}

% ============================================================
\section{Introduction}

Contemporary healthcare systems face an unprecedented demographic challenge: Eurostat projects that individuals aged 65 and above will comprise approximately 29\% of the EU population by 2050, exerting substantial pressure on institutional healthcare infrastructure~\cite{ec2021ageing}. The Internet of Medical Things (IoMT) -- interconnected wearable sensors, residential gateways, and cloud-hosted analytical platforms -- has emerged as the principal technological enabler of home-based elderly monitoring~\cite{razdan2022iomt}.

The data processed by IoMT systems constitutes a `special category' under GDPR Article~9~\cite{gdpr2016}, warranting heightened protection. Compromised IoMT systems pose direct patient safety risks -- falsified vital sign readings, suppressed emergency alerts, or manipulated device actuation -- extending beyond privacy harms to physical injury~\cite{kamalov2023iomt,enisa2025}.

Despite well-established cryptographic protocols, their application to resource-constrained IoMT for elderly care remains insufficiently addressed in existing literature~\cite{wani2024security}. Three principal impediments persist: (1) computational and energy constraints precluding heavyweight cryptography on sensor nodes; (2) architectural fragmentation where security controls are implemented piecemeal; and (3) regulatory-technical discontinuity between GDPR compliance requirements and engineering feasibility.

These impediments manifest as a tripartite research gap~\cite{wani2024security}: no prior validated work has simultaneously resolved edge-layer GDPR-compliant pseudonymisation, elderly-specific zero-interaction usability as a first-class architectural constraint, and systematic STRIDE-based validation of a complete unified architecture. This paper addresses all three through the SEG framework.

\subsection{Research Question}

How can a GDPR-compliant, edge-layer security architecture be designed and validated to simultaneously address the tripartite gap of cryptographic pseudonymisation, elderly-specific zero-interaction usability, and integrated STRIDE-based threat validation in resource-constrained IoMT deployments for remote elderly care?

% ============================================================
\section{Background and Related Work}

\subsection{IoMT Architecture}

IoMT systems for elderly care are organised into three functional layers~\cite{gatouillat2018iomt,lee2015iot}. The Perception Layer comprises resource-constrained wearable sensors -- pulse oximeters, thermometers, accelerometers -- characterised by limited processing capacity and small battery power supplies, which fundamentally constrain the range of deployable security mechanisms~\cite{algaradi2020survey}. The Network Layer performs mediation via a residential gateway, aggregating sensor data and managing secure upstream transmission; this layer is the primary focus of this work. The Application Layer encompasses cloud infrastructure and clinical interfaces.

\subsection{GDPR Requirements}

GDPR Article~25 mandates Privacy-by-Design~\cite{gdpr_art25}: data protection must be integrated from inception, not retrofitted post-deployment. Cavoukian's seven foundational principles~\cite{cavoukian2009pbد} -- proactive protection, privacy as default, embedded design, full functionality, end-to-end security, visibility, and user respect -- underpin this mandate and are directly operationalised in the SEG framework. Article~32 explicitly requires pseudonymisation and encryption as appropriate technical measures. Article~35 mandates a Data Protection Impact Assessment (DPIA) for high-risk processing, which health data monitoring constitutes.

\subsection{STRIDE Threat Model}

OWASP IoT security guidelines~\cite{owasp2024} identify authentication, access control, and encrypted communication as primary requirements for medical IoT deployments. STRIDE provides a systematic threat taxonomy across six categories -- Spoofing, Tampering, Repudiation, Information Disclosure, Denial of Service, and Elevation of Privilege -- as formalised by Shostack~\cite{shostack2014}. Each STRIDE category maps directly to specific architectural controls in the SEG framework.

\subsection{Related Work and Research Gap}

Table~\ref{tab:related} presents a structured comparison of the most closely related prior works. Rahmani et al.~\cite{rahmani2018} correctly identified the edge gateway as the appropriate security enforcement locus but did not address GDPR compliance or elderly usability. Koutli et al.~\cite{koutli2019} addressed GDPR partially but performed pseudonymisation at the cloud layer, leaving patient-identifiable data unprotected in transit. Rahman et al.~\cite{rahman2018blockchain} proposed a blockchain approach that provides audit immutability but imposes consensus latency incompatible with real-time emergency monitoring. Sun et al.~\cite{sun2019security} provide a comprehensive IoMT security survey but do not propose a validated architecture. Islam et al.~\cite{islam2025hybrid} propose a hybrid fog-edge architecture with health monitoring capability but without GDPR compliance mapping or elderly-specific design. Li et al.~\cite{li2023review} identify key security issues for precision health IoMT but stop short of an architectural solution addressing all three gap dimensions.

Recent work has reinforced the significance of the identified gaps. Dutta and Puthal~\cite{dutta2024ehealth} proposed a fuzzy logic and blockchain-enhanced IoMT-edge-cloud framework for eHealth in Society 5.0 (IEEE Access, 2024), demonstrating that elderly-specific IoMT architectures are an active research priority -- yet their framework does not address GDPR Art.~25 edge-layer pseudonymisation or zero-interaction usability. A CHI~2025 empirical study by Saka and Das~\cite{saka2025chi} found that only 13.64\% of elderly IoT users feel confident in existing privacy protections, with ``complex security settings'' generating distrust and anxiety -- directly corroborating the clinical necessity of the zero-interaction constraint adopted in the SEG. A systematic review by the same authors~\cite{saka2026sok}, accepted at ASIA CCS 2026, confirms that while more than 70\% of IoT studies implement encryption, fewer than 50\% address usability. No prior work addresses all three tripartite dimensions simultaneously.

\begin{table*}[!t]
\caption{Comparison of prior works against the tripartite research gap.}
\label{tab:related}
\centering
\renewcommand{\arraystretch}{1.3}
\begin{tabular}{|l|l|l|l|l|}
\hline
\textbf{Prior Work} & \textbf{Gap 1: Edge Pseudo.} & \textbf{Gap 2: Elderly UX} & \textbf{Gap 3: STRIDE} & \textbf{GDPR Art.25\&32} \\
\hline
Rahmani et al.~\cite{rahmani2018}   & Not implemented   & Not addressed & Not conducted & Not addressed \\
Koutli et al.~\cite{koutli2019}     & Cloud-layer only  & Not addressed & Not conducted & Partial       \\
Rahman et al.~\cite{rahman2018blockchain} & Not implemented & Not addressed & Not conducted & Not addressed \\
Sun et al.~\cite{sun2019security}   & Survey only       & Not addressed & Not conducted & Survey only   \\
Islam et al.~\cite{islam2025hybrid} & Not implemented   & Not addressed & Not conducted & Not addressed \\
Dutta \& Puthal~\cite{dutta2024ehealth} & Not implemented & Not addressed & Not conducted & Not addressed \\
\textbf{SEG (This Work)}            & \textbf{Edge-layer} & \textbf{Zero-interaction} & \textbf{Full 6-category} & \textbf{Art.25 \& 32} \\
\hline
\end{tabular}
\end{table*}

% ============================================================
\section{Methodology}

This research adopted Design Science Research Methodology (DSRM)~\cite{peffers2007dsrm}, appropriate for research whose primary scientific contribution is a novel artefact -- the SEG framework. DSRM accommodates rigorous artefact design and validation without requiring a live clinical deployment context. Three sequential phases structured the research.

Phase~1 (Problem Analysis) conducted a structured literature search across IEEE Xplore, ACM Digital Library, Scopus, and Google Scholar (2015--2025), identifying the tripartite gap and establishing clinical constraints for elderly care. Phase~2 (Design and Development) specified the SEG architecture, applied STRIDE threat modelling, and conducted an analytical DPIA structured in accordance with GDPR Article~35 and EDPB notification guidelines~\cite{edpb2022}. Phase~3 (Validation and Evaluation) validated the architecture through software-based PoC simulation, attack tree analysis, GDPR compliance mapping, and performance benchmarking against published literature. A feedback loop between Phases~2 and~3 ensured residual risk assessment before validation proceeded~\cite{peffers2007dsrm}.

% ============================================================
\section{SEG Framework Architecture}

The SEG framework is formally defined as: a three-tiered, edge-computing IoMT security architecture that enforces GDPR Article~25-compliant cryptographic pseudonymisation at the residential network boundary, implements elderly-specific zero-interaction usability constraints as binding architectural requirements, and has been validated against the complete STRIDE threat taxonomy within a unified design.

\subsection{System Architecture}

The architecture comprises three security zones. Zone~1 (Patient Zone) contains wearable sensors with no direct internet connectivity, connected to the gateway via wired protocols -- I2C for the MAX30100 pulse oximeter (address 0x57) and 1-Wire for the DS18B20 temperature sensor. Zone~2 (Residential Security Perimeter) houses the ESP32-WROOM-32 gateway, which enforces all security operations before data leaves the residential environment. Zone~3 (Cloud Zone) provides clinical access via RBAC and two-factor authentication.

\begin{figure}[!t]
\centering
\includegraphics[width=\columnwidth]{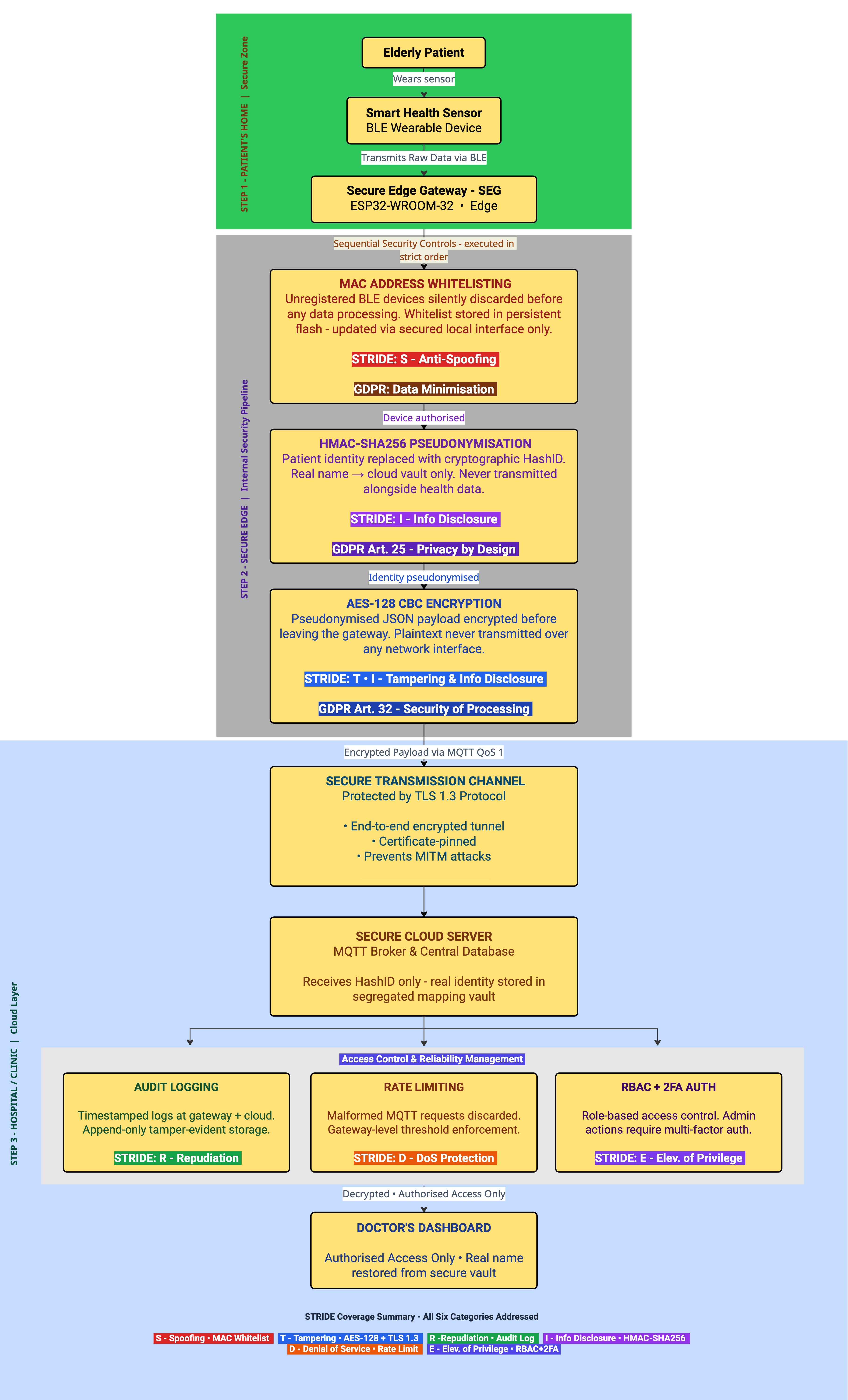}
\caption{High-level SEG security architecture across three zones.}
\label{fig:seg_arch}
\end{figure}

\subsection{Hardware Specification}

The SEG gateway is specified on the ESP32-WROOM-32D platform~\cite{espressif2023}: dual-core Xtensa LX6 at 240~MHz, 520~KB SRAM, hardware-accelerated AES cryptographic engine, and integrated dual-mode 802.11~b/g/n Wi-Fi + BLE~4.2 radio. This platform was selected for its hardware AES acceleration -- critical for energy-efficient encryption on a battery-adjacent device -- and native TLS~1.3 support via the ESP-TLS library.

\subsection{Data Processing Pipeline}

The pseudonymisation stage warrants particular attention as the primary privacy-enforcement step. Upon receipt of a validated sensor reading, the gateway retrieves the patient's registered identity and applies HMAC-SHA256 to derive a deterministic HashID -- for example, `a3f9c2b17e4d8f01' -- which replaces the patient identity field in the JSON payload before any further processing or transmission. Health data and patient identity are thereby separated at the earliest feasible point in the pipeline, consistent with GDPR Article~4(5) and ENISA pseudonymisation best practices~\cite{enisa2019pseudo}.

\subsection{Security Enforcement Mechanisms}

\textbf{ENISA-Compliant Pseudonymisation:} In compliance with GDPR Article~32(1)(a) and ENISA pseudonymisation best practices~\cite{enisa2019pseudo}, the gateway derives a pseudonymous HashID from the patient MAC address using HMAC-SHA256 with a gateway-resident secret key. The resulting 16-character hexadecimal digest (64-bit entropy) replaces the patient identity field before any network transmission. The gateway-side HMAC key is stored in persistent flash storage and updated only through a secured initial installation procedure conducted by a trained technician. This edge-layer pseudonymisation -- rather than cloud-layer -- is the primary architectural innovation: patient-identifiable data never traverses the public internet.

\textbf{AES-128-CBC Payload Encryption:} The pseudonymised JSON payload is encrypted using AES-128-CBC with a cryptographically random initialisation vector per transmission. AES-128 was selected over AES-256 based on empirical evidence that both provide equivalent security against all computationally feasible attacks, while AES-128 consumes measurably less energy per operation on resource-constrained hardware~\cite{hung2018aes}.

\textbf{TLS~1.3 + MAC Whitelisting:} All MQTT communication is encapsulated within a TLS~1.3 cryptographic tunnel~\cite{nist800207,nist80053}, providing transport-layer confidentiality and mutual authentication. The gateway cross-references the source device MAC address against a flash-stored whitelist prior to processing any packet; unregistered addresses are silently discarded, mitigating the STRIDE Spoofing threat.

\textbf{Elderly Zero-Interaction Constraint:} Consistent with empirical evidence that elderly users exhibit lower tolerance for complex technological interaction and heightened technology-induced anxiety~\cite{steele2009elderly}, the SEG is designed so that patients need perform no action beyond wearing the sensor devices. All authentication, pairing, encryption, and transmission occur automatically. This architectural choice simultaneously eliminates social engineering attack vectors including phishing, credential theft, and inadvertent misconfiguration.

% ============================================================
\section{Validation and Results}

\subsection{STRIDE Threat Analysis}

Table~\ref{tab:stride} presents the detailed technical security risk assessment of the SEG architecture, mapping each STRIDE threat category to a specific attack scenario, the target component, the qualitative risk severity rating assessed using CVSS~v3.1 categories, and the architectural countermeasure implemented to mitigate each identified threat~\cite{shostack2014}.

\begin{table*}[!t]
\caption{Technical security risk assessment and countermeasures. Risk levels assessed using CVSS~v3.1 qualitative categories.}
\label{tab:stride}
\centering
\renewcommand{\arraystretch}{1.4}
\begin{tabular}{|p{1.5cm}|p{3.2cm}|p{2.2cm}|p{1.2cm}|p{5.5cm}|}
\hline
\textbf{STRIDE Threat} & \textbf{Attack Scenario} & \textbf{Target Component} & \textbf{Risk Level} & \textbf{Architectural Countermeasure} \\
\hline
Spoofing &
Attacker deploys rogue sensor with fabricated health data to inject falsified readings &
BLE / Gateway Interface &
High &
MAC address whitelisting: unregistered devices silently discarded prior to data processing \\
\hline
Tampering &
Man-in-the-middle interception and modification of health data values in transit &
Network Data Stream &
Critical &
HMAC integrity verification: modified payloads detected and rejected; AES-128 encryption prevents plaintext manipulation \\
\hline
Repudiation &
Authorised user performs administrative actions and subsequently denies them, exploiting absence of audit trail &
Cloud Database / Gateway &
Medium &
Comprehensive timestamped audit logging at both gateway and cloud layers; log integrity protected via append-only storage \\
\hline
Information Disclosure &
Attacker captures transmitted MQTT packets or accesses cloud storage to read patient identity and health records &
Storage / Network &
Critical &
Cryptographic pseudonymisation segregates identity from health data; AES-128 renders captured payloads computationally undecipherable \\
\hline
Denial of Service$^\dagger$ &
Attacker floods gateway with malformed MQTT connection requests to exhaust processing capacity and suppress alerts &
Gateway &
High &
Connection rate limiting: threshold-exceeded connections rejected; edge processing maintains alert functionality independently of internet availability \\
\hline
Elevation of Privilege &
Attacker exploits firmware vulnerability to acquire administrative gateway control, enabling full system compromise &
Gateway OS / Firmware &
High &
Principle of least privilege enforced; administrative interface restricted to local physical access; regular firmware integrity verification \\
\hline
\end{tabular}

\smallskip
\footnotesize $^\dagger$Large-scale volumetric DDoS attacks must be mitigated at the upstream ISP network level; the edge-based rate limiting implemented here is designed specifically to address local DoS attempts~\cite{nist800207,nist80053}.
\end{table*}

The attack tree analysis demonstrates that successful compromise of the Edge Gateway requires an attacker to overcome multiple independent defensive layers, consistent with the Defence-in-Depth security principle~\cite{nist800207,nist80053}. No single point of failure was identified across all six STRIDE categories within the scope of this simulation-based analysis.

\subsection{GDPR Compliance Mapping}

Table~\ref{tab:gdpr} presents the compliance mapping of SEG architectural features against GDPR Articles~25 and~32 obligations. All nine identified regulatory obligations are satisfied.

\begin{table*}[!t]
\caption{GDPR compliance mapping -- all nine obligations architecturally addressed.}
\label{tab:gdpr}
\centering
\renewcommand{\arraystretch}{1.4}
\begin{tabular}{|p{3.2cm}|p{2.2cm}|p{7.0cm}|p{2.0cm}|}
\hline
\textbf{GDPR Obligation} & \textbf{Article} & \textbf{SEG Implementation} & \textbf{Status} \\
\hline
Data Protection by Design & Art.~25(1) & Security integrated from initial design phase & Architecturally addressed \\
\hline
Data Minimisation & Art.~25(1) & Three parameters only: HR, SpO$_2$, Temperature & Architecturally addressed \\
\hline
Privacy by Default & Art.~25(2) & Maximum protection without user configuration & Architecturally addressed \\
\hline
Pseudonymisation & Art.~32(1)(a) & HMAC-SHA256 HashID at residential edge boundary & Architecturally addressed \\
\hline
Encryption & Art.~32(1)(a) & AES-128-CBC application + TLS~1.3 transport & Architecturally addressed \\
\hline
Ongoing Confidentiality & Art.~32(1)(b) & MAC whitelist + dual-layer encryption & Architecturally addressed \\
\hline
Ongoing Integrity & Art.~32(1)(b) & HMAC-SHA256 per-packet integrity verification & Architecturally addressed \\
\hline
Ongoing Availability & Art.~32(1)(b) & Rate limiting; edge-local alerting & Architecturally addressed \\
\hline
Audit Logging & Art.~32(1)(c) & Timestamped tamper-evident event logs & Architecturally addressed \\
\hline
\end{tabular}
\end{table*}

\subsection{Performance Analysis}

\textit{Note on evaluation methodology:} All performance figures cited in this section are derived exclusively from published peer-reviewed empirical benchmarks, as noted per citation. Physical hardware measurement falls outside the scope of this simulation-based study and is identified as Priority~1 in the future work roadmap.

\textbf{Protocol energy:} MQTT's persistent TCP connection eliminates per-request handshake overhead. Yassein et al.~\cite{yassein2016mqtt} document MQTT's 2-byte binary fixed header versus HTTP's 200--800-byte text header per transaction. Empirical measurement by Jara Ochoa et al.~\cite{jara2023mqtt} demonstrates MQTT consumes 6--8\% less energy than HTTP under comparable IoT deployment conditions -- a meaningful difference for continuous, battery-adjacent health monitoring.

\textbf{Latency:} Edge computing consistently achieves sub-50~ms response latencies for local gateway processing -- reported as low as 11.5 ms in comparable edge gateway deployments~\cite{pace2018edge} and confirmed across IoMT-specific deployments~\cite{verma2018fog,dong2020edge} -- compared to 200--700~ms for cloud-only processing under normal conditions. The SEG's edge-local emergency alerting provides sub-50~ms propagation independent of cloud connectivity -- a clinically critical property for acute cardiac or respiratory event detection.

\textbf{Software-based PoC validation:} A Python-based simulation of the complete SEG processing pipeline -- MAC whitelisting $\rightarrow$ HMAC-SHA256 pseudonymisation $\rightarrow$ AES-128-CBC encryption $\rightarrow$ JSON serialisation $\rightarrow$ MQTT formatting -- confirmed the operational feasibility of the SEG pipeline, with each processing stage executing correctly in sequence, as shown in the simulation output log.

% ============================================================
\section{Discussion}

\subsection{Privacy-by-Design Implementation Challenges}

Implementing GDPR Article~25 on resource-constrained microcontroller hardware reveals fundamental tensions between regulatory ideality and engineering reality. The data minimisation principle requires upfront clinical judgements about which health parameters are `strictly necessary'. The SEG resolves this by restricting collection to three clinically universal elderly monitoring parameters: heart rate, peripheral oxygen saturation, and body temperature. Pseudonymisation at the gateway layer -- rather than a cloud service -- ensures the Article~4(5) identity-data separation is enforced at the earliest feasible point in the data pipeline, providing stronger privacy guarantees than any cloud-layer pseudonymisation approach.

\subsection{Zero-Interaction Usability as Clinical Necessity}

A security architecture that imposes cognitive or interaction burden on elderly users ultimately fails its clinical purpose. Saka and Das~\cite{saka2025chi} found empirically that only 13.64\% of elderly IoT users feel confident in existing protections, with complex security settings generating distrust and anxiety -- recommending `privacy and security by design' with simplified interfaces. A systematic review by the same authors~\cite{saka2026sok} confirms this gap persists across the most recent literature: fewer than 50\% of IoT studies address usability. By eliminating all patient-facing authentication -- no passwords, no manual pairing, no active inputs -- the SEG simultaneously removes the entire class of social engineering attack vectors. This operationalises the zero-interaction constraint recommended by~\cite{saka2025chi} as a binding architectural requirement rather than a post-hoc usability concern, consistent with Steele et al.'s~\cite{steele2009elderly} earlier findings on elderly wireless sensor acceptance.

\subsection{Comparison with Existing Systems}

The SEG framework demonstrates improvements across all evaluated dimensions relative to conventional cloud-based IoMT deployments: edge processing achieves sub-50~ms latency (as low as 11.5 ms in comparable deployments~\cite{pace2018edge}) versus 200--700~ms cloud-only~\cite{verma2018fog,dong2020edge}; MQTT reduces energy consumption by 6--8\% versus HTTP~\cite{yassein2016mqtt,jara2023mqtt}; and pseudonymisation is applied before transmission. The central finding challenges the common assumption of an inherent trade-off between security and operational efficiency: in the SEG architecture, security controls and performance are mutually reinforcing objectives.

\subsection{Limitations}

This study acknowledges the following limitations, each defining a future work priority. The evaluation methodology is simulation-based rather than empirical -- physical hardware energy consumption and latency were not directly measured. Synthetic rather than genuine patient data was employed. The regulatory analysis is confined to the GDPR framework applicable within the EEA. Physical hardware deployment and empirical HCI testing with real elderly patients fall outside the current study scope. MAC whitelisting does not address physical sensor theft scenarios.

% ============================================================
\section{Conclusions}

This paper presented the Secure Edge Gateway (SEG) framework, addressing the tripartite research gap in IoMT security for elderly care through three original contributions.

\textbf{Contribution 1:} A software-simulation-validated IoMT 
architecture implementing GDPR Article 25-compliant HMAC-SHA256 cryptographic pseudonymisation at the residential edge boundary before any network transmission, ensuring patient-identifiable data never traverses the public internet.

\textbf{Contribution 2:} A zero-interaction security model implementing elderly-specific usability constraints as first-class architectural requirements, eliminating all patient-facing authentication and simultaneously removing social engineering attack surfaces.

\textbf{Contribution 3:}  A systematic STRIDE-based threat validation -- across all six categories -- of a complete, unified IoMT architecture incorporating both edge pseudonymisation and 
zero-interaction usability constraints.

The central finding -- that GDPR compliance and operational efficiency are complementary rather than competing objectives in resource-constrained IoMT elderly care -- has significant implications for practitioners designing European market healthcare monitoring systems.

Future work will pursue five phases: physical ESP32-WROOM-32D hardware deployment with empirical energy and latency measurement; TinyML-based personalised anomaly detection on the edge device~\cite{sanchez2020tinyml}; lightweight blockchain audit integrity; IPv6 and 5G scalability; and X.509 cryptographic device attestation to address the physical theft limitation.

% ============================================================
\section*{Acknowledgment}
The authors would like to thank Principal Lecturer P\"{a}ivi Haho and Senior Lecturer Sakari Lukkarinen of Metropolia University of Applied Sciences for their valuable guidance and feedback during the preparation of this work.

% ============================================================

\end{document}